\documentclass [11pt,twoside]{article}\usepackage{mmap}
\usepackage[margin=1in,centering]{geometry}\usepackage[unicode]{hyperref}
\newcommand\aut {Leonid A.~Levin}\newcommand\ttl {Taxation and Valuation}
\newcommand\hreff[1] {{\footnotesize\href{http://#1}{http://#1}}}

\begin{document} \date{}\title {\vspace*{-6pc}\ttl\thanks
 {{\em Tax Notes \bf 164}(7). \href{https://www.taxnotes.com/tax-notes-federal/tax-policy/taxation-and-valuation/2019/08/12/29rhn}
{www.taxnotes.com/tax-notes-federal/tax-policy/taxation-and-valuation/2019/08/12/29rhn}}
 \vspace*{-1pc}} \author {\aut\thanks {Boston University, CS dept.,
 111 Cummington Mall, Boston, MA; Home page: \hreff{www.cs.bu.edu/fac/Lnd/}}}
 \maketitle\renewcommand\abstractname {Summary}\vspace*{-2pc}

\begin{abstract}
 The greatest harm from highway robbers lies not in seized wallets but in
inhibited travel. Similarly, incentives for tax-reducing strategies put much
sand in the wheels of the economy. Demands to replace our monumental tax code
with a simple, graceful one that does not distort economic incentives heat up
periodically in political debate, but such dreams never materialize.

A {\bf fundamental} obstacle, not yet well understood in the economic
literature, is the impossibility of objectively evaluating the tax base --
assets, income, etc. One can see this even in toy examples, say, trying to
assess the value of a position in chess: great masters' assessments will all
differ. (Here computer theory can add an insight not provided by classical
economics tools.)

A way around is to avoid evaluations by expressing the tax in natural units,
not in cash. For publicly traded corporations, these could be corporate shares.
I discuss a simple (postcard-sized in {\bf all} details) corporate tax system
that avoids {\bf any} distortion of incentives. (Tax tools {\em meant\/}
to influence corporate policies should be set as explicit separate taxes
or credits, open to public scrutiny, not hidden between lines of an
incomprehensible tax code.) Roughly, the~system is to periodically
take a $t{\cdot}i$ fraction of shares to auction, where $t$ is the tax rate,
$i$ is the interest rate. It replaces all income taxes on publicly traded
corporations, their subsidiaries, and shareholders.

\hspace*{-11pt}{\footnotesize
 (One way to view this is that tax on income invested in the
 public sector is deferred -- treated~as~interest-bearing debt.
 This obviates the need to determine which part of the corporate value is
untaxed income: All of it is, if investments into the public sector are pretax,
divestments taxed, or, as in our equivalent but simpler case, vice versa.)}
 The interest rate is defined via specially designed bonds, so that
the whole system can be shown {\bf precisely} equivalent to a flat tax
on {\bf investment return}. Note that taxing the return {\bf directly}
is impossible: it would invite manipulation of stock market~prices.

The main feature is that nothing corporations and investors do can change
their tax ($t{\cdot}i$ fraction of shares), so they would do business
exactly the {\bf same way} they would {\bf without~taxes.} \end{abstract}

\subsection* {Cash Taxes Cannot Avoid Distortion of Incentives}

Taxes have major costs beyond the revenues they collect: deadweight loss
from distorted incentives, costs of compliance, enforcement costs, etc.
(The report of the 2005 President's Advisory Panel on Federal Tax Reform
mentions a trillion dollar annual waste.) Countless attempts to alleviate these
effects invariably just shifted the distortions from one place to another.

Here is the key observation: Modern economics, based on classical game theory,
assumes rational optimization of some consistent, legally definable values such
as assets or their growth (i.e.\@ income). This, however, fails to recognize
the infeasibility of consistent valuations and other types of optimal,
rational behavior in many games (much more so in real life).\footnote
 {Economists do now recognize, besides grain, land, and coal, the relevance
of another commodity: information.\\ I doubt all fully realize how subtle
this concept is, but even a cursory attention to it has brought progress.\\
Yet one more factor -- intelligence -- needs acknowledging. Even with full
and perfect information,\\ the IRS couldn't match all taxpayers in intelligence,
and thus in the ability to evaluate assets.\\ Lacking such ability,
it acts like a bull in a china shop, vandalizing our economic life.}

For instance, to play chess the first approach that comes to mind is to
understand how to compute the positions' value, and to choose each move
to maximize it. The value must be consistent across a move, i.e.\@
agree with the best value of the next position one move can achieve.
Indeed, each position does have such a consistent $\{\pm1,0\}$ value:
one side has a winning strategy or both have a draw. Just keep moving
to positions of the same value. What a silly way to pass the time!

What keeps it fun is the {\bf exponential} computation any such strategy has
been proven to require! I argue that any feasible legal definition of the tax
base value will be inconsistent with taxpayers' motives and thus distortive.
(Taxing any feasibly defined gain in chess positions would change the game
entirely.) There are, however, unusual yet neat, sound, and practical ways
around.

\subsection* {A Corporate Tax Code on a Postcard}

First, the market-clearing interest rate \fbox {$i$} is set via TIPS bonds
designed so that either side -- Treasury and a publicly traded corporation
({\bf PTC}) -- could unilaterally get any desired bond~exposure at that rate.
Treasury must absorb all differences between supply and demand by buying
those bonds back at the (inflation-adjusted) purchase price or issuing more.
But it can change $i$ at will (with due notice so customers
can buy or sell bonds before the new rate takes effect).
So it controls the supply and demand to keep its desired bond exposure, too.
 The PTC tax rate \fbox {$t$} is just set by the law.
 {\footnote {Some wishful thinking: The US Constitution requires fair
  compensation for private property taken for public use. This seems to
  imply spending taxes to fairly benefit the taxpayers, e.g., giving them a
  tax-weighted say in approving public spending levels. Then they would do
  a better job than Congress in setting tax rates optimal for growth.}}
 It should agree with the effective private sector rate, to
 keep the net capital flow between the PTC's and the private
 sectors tax revenue neutral. Now, the full PTC tax code:

\vfil\noindent\fbox{\parbox{\dimexpr\hsize-10pt}
 {At regular dates (also on in-dividend dates), PTCs give the IRS to auction
 a $t{\cdot}i$ fraction of shares. They buy back shares for this or issue more.
 Subsidiaries, wholly owned by PTCs, pay no tax.\\ A {\em partial subsidiary},
 owned in part by PTCs, pays the respective part of its tax to its\\ PTC owners,
 not to the IRS, or have some equivalent protection from double-taxation.

 \vspace{6pt} Not voting bond-like securities can be taxed similarly if
 they are tradable in fractions.\\ But a simpler equivalent is to tax their
 proceeds at a $(1{-}e^{-st})$ rate where $s$ is the sum\\ of daily interest
 rates for the entire time the security was held outside the PTC sector.

\vspace{6pt} Going public turns prior shares' cost bases tax deductible
 to shareholders but triggers a {\em conversion tax\/}: giving the
 IRS (to auction) {\em options\/} to buy a fraction $t$ of shares
 at the {\em strike\/} price totaling all corporate income tax to date.
 (Similar ``strike price credits'' can be used later for other taxes e.g.,
 foreign taxes under US treaties.) Reconverting to private, a company
 can establish its shares' cost basis, $b$, by giving the IRS {\em put\/}
 options for a fraction $t$ of its shares at strike price $b$.}}\vfil

This code replaces corporate income tax on PTCs, and taxes on their
shareholders' dividends and capital gains. It distorts no incentives: boosting
post- and pre-tax values is exactly the same. Its enforcement and compliance
costs are minimal. It requires no complicated regulations, except those
unrelated to taxes -- say, protecting minority shareholders. Impossibility to
hide or delay liability lowers tax rates. A steady trickle of auctioned shares
may even have some stabilizing effect on the stock market. This Equity Interest
Tax ({\bf EIT}) emulates the $t$-rate tax on (real) return\footnote
 {which cannot be taxed in cash based on stock prices,
  lest firms manipulate the market, undermining its integrity.}:
  market pressures would keep the variable $i$ close to return rates.

 The precise match with tax on return is clearer via EIT's equivalent but a bit
more cumbersome variant: EIT*. It differs from EIT like IRAs from Roth IRAs:
all investments (from private sector) are income tax deductible
and divestments are income-taxed (both at rate\footnote
 {Personal tax is progressive. But it can be viewed as a flat rate $t$ applied
after deducting typical living expenses which grow sublinearly with earnings.}
$t$). Then the entire stock market capitalization $V$ would be untaxed income,
the deferred tax $t{\cdot}V$ on it -- an enormous loan from Treasury. To
finance it Treasury can sell bonds and pay interest $i{\cdot}t{\cdot}V$ on
them, compensated by EIT* (proceeds from the $i{\cdot}t$ flow of auctioned
shares). Similarly, any company can spend this loan (deferred tax) on bonds,
the interests on which would compensate its EIT*. The net expense is
$t{\cdot}r$, to update the bond portfolio as the company's worth grows by its
return~$r$.

\vfil\subsection* {Just One Issue in a Broader Scope}

The tools discussed work only for the PTC sector.
The point was that the failure of all persistent tax reform efforts had a
cause that, while fundamental, can be circumvented in important cases.\\
The above tools cannot add grace to taxes on closely held business or personal
earnings.\\ Yet those, too, have aspects that can benefit from reforms.
Some widely discussed examples:

{\bf Dividends and capital gains} taxes have low rates but apply largely
to income already taxed at the corporate level. This is widely criticized.
Making dividends (paid from taxed income) tax free and allowing companies
to deduct capital losses (up to per-share taxed income) on share repurchase
would be more consistent than lower tax rates on dividends, capital gains,
and corporate income.

{\bf Gifts/estates} also should avoid double taxation.
Stock market income and gifts received can be excluded, because they should
have been fully taxed already. The rest, instead of a large standard deduction,
can get a tax credit for all income taxes the donor ever paid.

{\bf Taxes on medical expenses} penalize deductibles in medical insurance.
Needlessly low deductibles make one careless with expenses\footnote
 {The insurance effect is diffusion of responsibility.
  This agrees with the general liberal ideology which has a reason:
  Since society absorbs much of the rewards of one's success,
  it should also absorb much of the pains of one's failure.
  Otherwise, people would have a suboptimal risk tolerance.
  The conservative counterargument seems to be:\\
  ``While three lefts make a right, two wrongs do not.'' :-).
  General tax policies should be neutral on such issues.}
 which is widely blamed for skyrocketing medical costs.
 To rectify this tax-induced distortion, insurance-approved
 out-of-pocket medical costs (deductibles, co-payments, coinsurance, etc.)
 could be reported by insurance and be fully tax deductible.

{\bf Many other} concerns and ideas would, of course,
resurface with the tax reform drive\\ heating up again.
E.g., taxing residential rent expense depresses population mobility.\\ (As they
say, ``When a tenant marries the landlord, the national income shrinks.'' :-)

The topic of the publicly traded sector is just one of a great
many, but it\\ is a large one, assuring that at least some
significant improvements are achievable.

\vfill\end{document}